\documentclass[a4paper,twoside,twocolumn]{article}
\pdfoutput=1

\newif\iftr 
\trtrue     
\newif\ifccs   
\ccstrue      

\usepackage{ifpdf}  
\ifpdf
    \usepackage[pdftex]{graphicx}
    \usepackage[colorlinks,linkcolor=blue,citecolor=blue,urlcolor=blue]{hyperref}
    \DeclareGraphicsExtensions{.pdf}
\else
    \usepackage{graphicx}
    \usepackage[hypertex]{hyperref}  
\fi

\usepackage{todonotes} 

\usepackage{
url       %
,fancyhdr 
,lastpage 
,enumerate 
,amsmath  
,amsfonts 
,amssymb  
, enumitem 
, array 
}
\usepackage[hang]{footmisc} 
\usepackage[noindent, arraystretch, fullpage]{setlengths} 
\usepackage{
own         
}

\graphicspath{{images/}}

\newcommand*{\metaauthori}{Bob Briscoe}

\newcommand*{\metashorttitle}{The Native AQM for L4S Traffic}
\newcommand*{\metatitle}{{\large{Technical Report}}\\The Native AQM for L4S Traffic}
\newcommand*{\metano}{TR-BB-2017-002}
\newcommand*{\metakeywords}{Data Communication, Networks, Internet, Control, Congestion Control, Quality of Service, Performance, Latency, Responsiveness, Dynamics, Algorithm, Standards, Active Queue Management, AQM, Sojourn time, Virtual queue, Explicit Congestion Notification, ECN, Gradient Marking, Ramp}

\newcommand*{\metamaili}{\href{mailto:research@bobbriscoe.net}{research@bobbriscoe.net}}

\newcommand*{\metaaddress}{}

\newcommand*{\metaversion}{02}
\newcommand*{\metadate}{15 Apr 2019}

\hypersetup{                       
     pdfauthor = {\metaauthori%
     },
     pdftitle = {\metashorttitle},
     pdfsubject = {},
     pdfkeywords = {\metakeywords}
}%

\title{\metatitle}%
\author{\metaauthori%
\thanks{\metamaili, %
\metaaddress}%
}
\date{\metadate}%

\fancypagestyle{first}{%
\fancyhead[LO,RE]{}%
\fancyhead[LE,RO]{}%
\fancyhead[C]{}%
}%

\begin{document}
\bibliographystyle{alpha}%


\maketitle%
\thispagestyle{first}

\begin{abstract}
{\small\noindent%
This memo focuses solely on the native AQM of Low Latency Low Loss Scalable throughput (L4S) traffic and proposes various improvements to the original step design. One motivation for DCTCP to use simple step marking was that it was possible to deploy it by merely configuring the RED implementations in existing hardware, albeit in an unexpected configuration by setting parameters to degenerate values. However, there is no longer any imperative to stick with the original DCTCP step-function design, because changes will be needed to implement the DualQ Coupled AQM anyway, and the requirements for L4S congestion controls are not yet set in stone either. This paper proposes gradient (ramp) marking and a virtual (a.k.a.\ phantom) queue. It provides a way to implement virtual queuing delay (a.k.a.\ virtual sojourn time) and scaled virtual sojourn time.}      
\end{abstract}
\ifccs{}%
%
%
\subsection*{CCS Concepts}
\textbf{\textbullet Networks} \(\to\) \textbf{Cross-layer protocols; Network algorithms; Network dynamics;}
\subsubsection*{Keywords}
\metakeywords
\fi{}%

\section{Introduction}\label{l4saqmtr_intro}

The DualQ Coupled AQM~\cite{DeSchepper15b:DCttH_TR, Briscoe15e:DualQ-Coupled-AQM_ID} was proposed as a way to provide a Low Latency Low Loss and Scalable throughput (L4S) service that would be usable for all Internet traffic, while doing no harm to existing non-L4S (`Classic') traffic. 

To preserve the low latency service, L4S traffic has to use a so-called `Scalable' congestion control. DCTCP~\cite{Alizadeh10:DCTCP} is a good example for elastic traffic, but it needs some minor safety and performance modifications for use on the public Internet; termed the `TCP Prague `requirements~\cite{Briscoe15f:ecn-l4s-id_ID}. Other scalable congestion controls are being developed, for instance for real-time applications.

\begin{figure}[h]
	\centering
	\includegraphics[width=\columnwidth]{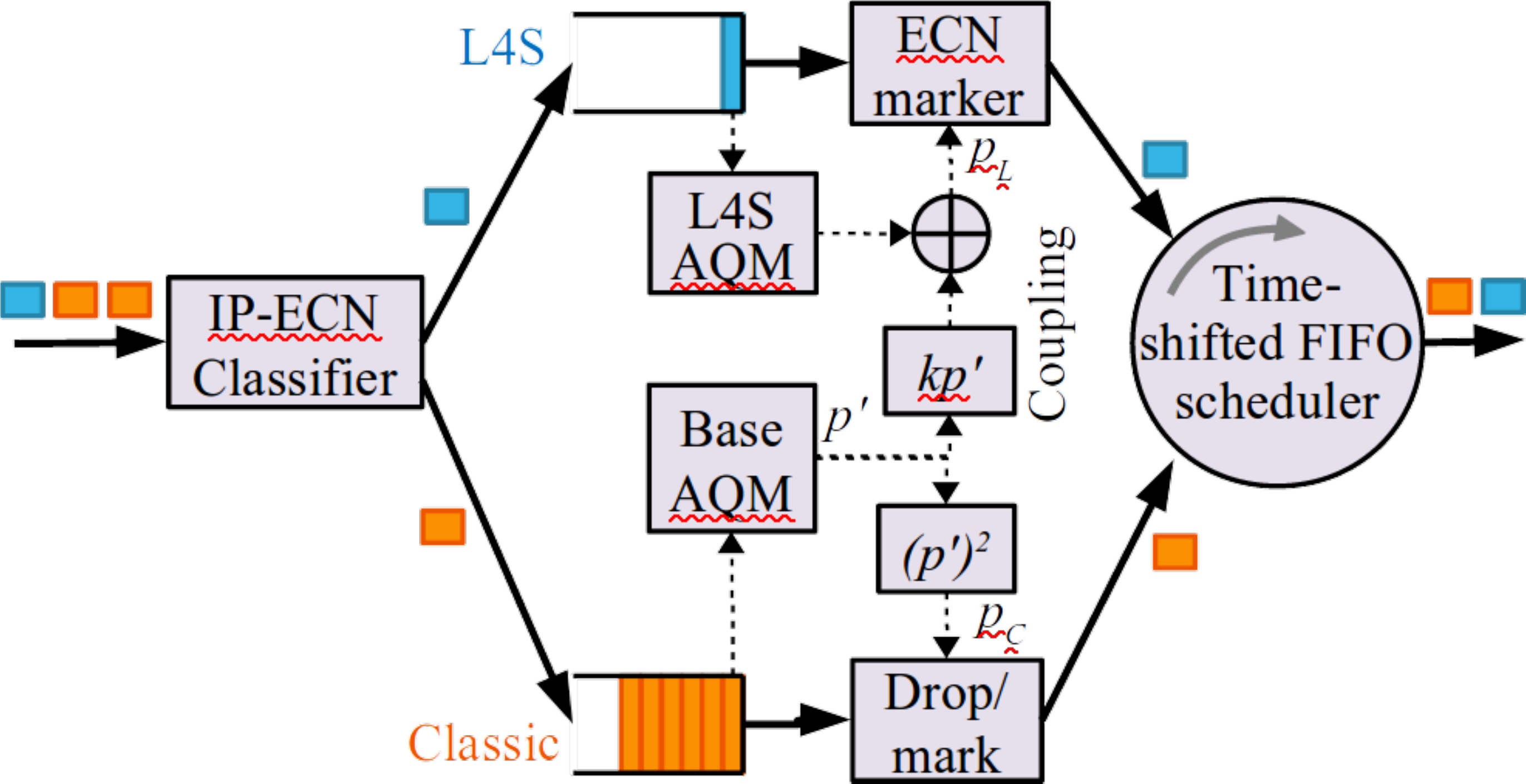}
	\caption{DualQ Coupled AQM}\label{fig:aqm-dualq-coupled}
\end{figure}

The Dual AQM is shown in \autoref{fig:aqm-dualq-coupled} where the two queues for L4S and Classic traffic can be seen. Each queue is regulated by its own native AQM, but the AQM for Classic traffic serves as a Base AQM both for Classic traffic and for to couple the marking of L4S traffic to that of Classic. 

Coupled marking is necessary for coexistence between L4S and Classic traffic, without the former starving the latter. As shown in the figure, marking from the base AQM is applied linearly to L4S traffic, but squared before applying to Classic traffic. The squaring counterbalances the square root in the Classic TCP rate equation, so that each flow ends up using roughly the same share of the total link capacity, irrespective of whether it is a Classic or a L4S flow (see the above references for details).

The DualQ Coupled AQM includes a native L4S AQM to regulate L4S traffic on its own. When both L4S and Classic traffic are present, L4S traffic is marked if either the base AQM or the native L4S AQM decide to mark it, which is shown as a logical OR symbol (\(\oplus\)) in the figure.

When the DualQ Coupled AQM was first proposed, the L4S AQM was implemented and evaluated as a simple step ECN marking function based on the instantaneous queue length, given this was already proven for DCTCP. Nonetheless, subsequent work has highlighted some issues with this simple AQM. 

This memo focuses solely on the native L4S AQM and proposes various improvements to the original step design. One motivation for DCTCP to use simple step marking was that it was possible to deploy it by merely configuring the RED implementations in existing hardware, albeit in an unexpected configuration by setting parameters to degenerate values. However, there is no longer any imperative to stick with the original DCTCP step-function design, because changes will be needed to implement the DualQ Coupled AQM anyway, and the requirements for L4S congestion controls are not yet set in stone either.

Briefly, the solutions explored in this memo are:
\begin{itemize}
	\item Using a virtual queue
	\item Gradient marking, that is, using a (steep) ramp or convex function rather than an on-off step
\end{itemize}

An introductions to the issues that each of these solutions addresses is given at the start of each individual section about the solution. These solutions can each be applied independently or in any combination, including using all of them.

\subsection{The Current Native L4S AQM}\label{l4saqmtr_current}

The current reference software implementation of the DualQ Coupled AQM for Linux (specified as pseudocode in Appendix A of \cite{Briscoe15e:DualQ-Coupled-AQM_ID}) has already been changed from the original DCTCP-based design. This is because configuration of a step threshold in units of bytes depends on the drain rate, but the drain rate of the L4S queue continually varies depending on the proportion of Classic traffic competing for the link. Evaluations have proved that the original queue-length-based measurement led to increased L4S queuing delay when Classic traffic was present (because a certain number of bytes in a queue clearly takes longer to drain when the drain rate is reduced).

It has subsequently been proposed by others as well that the DCTCP step function ought to be configured in time units---for the similar reason that the drain rate might vary if the queue is part of a larger scheduling hierarchy~\cite{Nichols12:CoDel, Bai16:MQ-ECN, Bai16:ECN_GPS}.

Therefore, the current L4S native AQM uses the sojourn time of each packet to measure the L4S queue in units of time, not bytes. It marks a packet if the sojourn time is greater than a fixed time threshold (1\,ms for the public Internet) unless the queue length (in bytes) is less than 2 MTU. The latter condition is necessary to prevent excessive marking in cases where the drain rate is very low, so that a 1\,ms threshold would represent less than 2 full-sized packets. Otherwise the resulting marking would severely under-utilize the link.


\section{DualQ Coupled Marking with a Virtual Queue}\label{l4saqmtr_vq}

\subsection{Virtual Queues}

A virtual queue (VQ)~\cite{Courcou96:Vq} is a mechanism used to model how long the queue would be if the packets arriving at the real queue were served by a link with a virtual capacity lower than the real link, typically only slightly lower. The length of the VQ can be used to drive an AQM algorithm. 

The AQM may be intended for a particular type of traffic, e.g.\ to aid admission control of inelastic traffic as in pre-congestion notification (PCN)~\cite{Eardley09:PCN-mark}. However, the load metric from a VQ can usually replace the same metric from the real queue in any AQM intended for a mix of traffic types~\cite{Kunniyur01:AVQ}. 

A VQ-based AQM emits congestion signals (drops or ECN marks) into the forwarded data, which are fed back to all the congestion controls loading the real link as usual. This fools the sources into aiming to fill only the virtual link capacity. In this way, the real link should remain (slightly) under-utilized and its queue should remain very small.

Because a VQ does not introduce appreciable queuing delay, mechanisms like TCP's ACK clock no longer automatically pace out new packets, so TCP continually introduces bursts into the network unless the source applies its own pacing~\cite{Alizadeh12:HULL}, either continually or at least during slow-start (as is currently already done in the Linux TCP stack). 

\begin{figure}[h]
	\centering
	\includegraphics[width=\columnwidth]{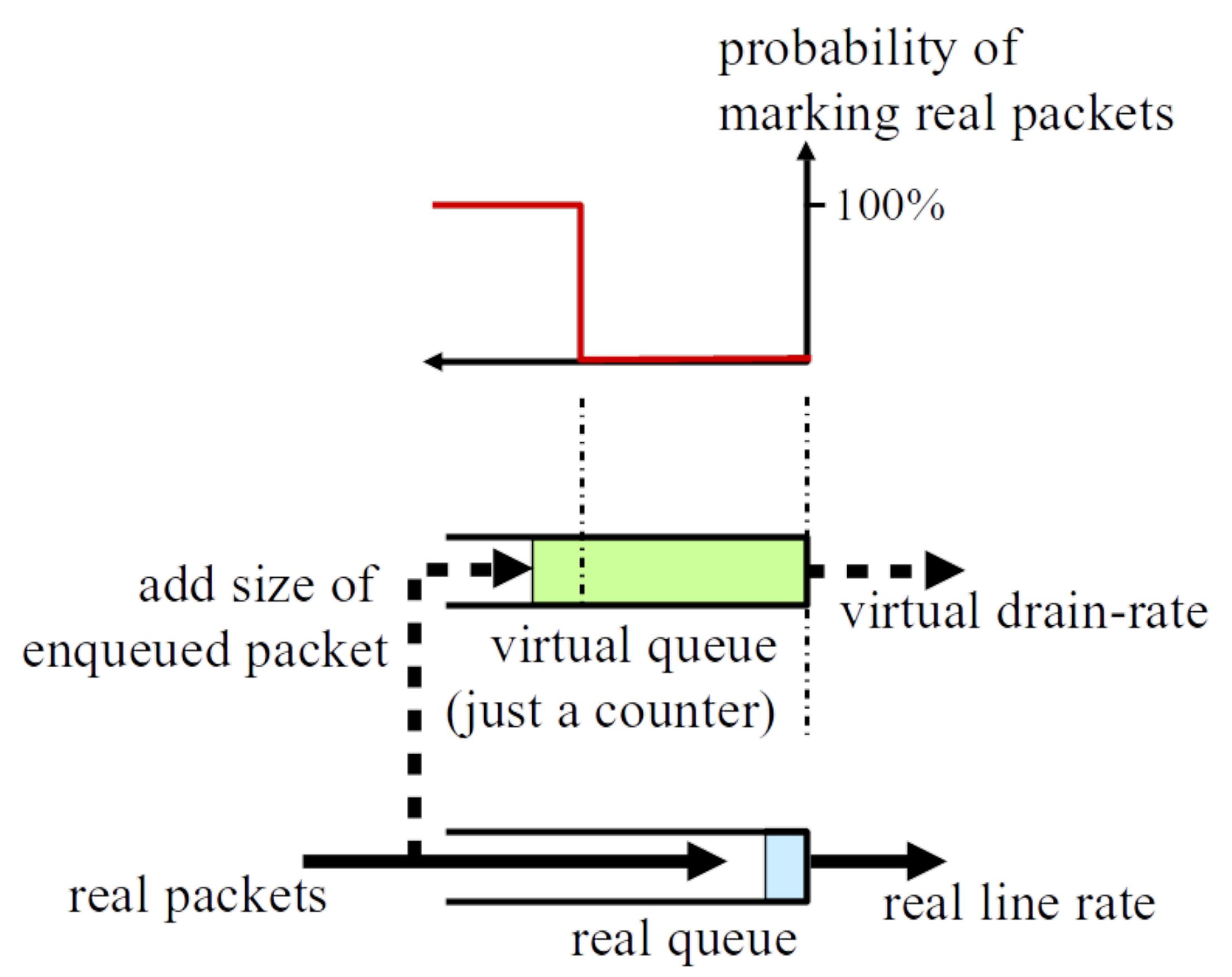}
	\caption{A Virtual Queue}\label{fig:vq}
\end{figure}

As \autoref{fig:vq} shows, a VQ is actually just a number representing the length of the VQ (in bytes). When a packet is enqueued into the real queue, its size is added to the VQ. When a packet is dequeued from the real queue, slightly less than its size is subtracted from the VQ (with a floor of zero, as for a real queue). Other implementation techniques are possible. For instance, for a fixed rate link, the virtual queue could be drained at regular intervals, at slightly less than the real link rate.

We will define the ratio between the virtual and real link rates as \((1-\varepsilon)\), where \(\varepsilon\) is a small positive fraction. For implementation efficiency it should preferably be a power of 2, for example \(1/64\). If the congestion control system was perfect, this would drive the link utilization to \(63/64 \approx 98.4\%\).

\subsection{VQ as the Native L4S AQM}\label{l4saqmtr_vq-l4s}

Na\"{\i}vely, it might be thought that using a virtual queue only for the L4S AQM might upset the balance between Classic and L4S capacity shares.

However, as long as \(\varepsilon\ll1\), the difference should be negligible---certainly a lot smaller than existing differences in capacity shares due to other factors, such as differences in congestion control design and implementation as well as RTT-dependence. This assertion will need to be tested, but the intuition is as follows. 

Consider a case where \(n_L\) long-running L4S flows are present. If \(n_C\) long-running Classic flows are present, the coupling adds an appropriate amount of marking to L4S traffic, as if \(n_C\) L4S flows were mixed with the \(n_L\) actual L4S flows. The congestion controllers of each L4S flow also drive the native L4S AQM to the point where its marking probability reduces their congestion windows to fit into the total link capacity with the same congestion window for each flow (Classic and L4S). Equivalently, one can consider that the coupled marking causes the L4S flows not to utilize \(n_C/(n_L+n_C)\) of the capacity, while the level of L4S marking behaves as if the \(n_L\) L4S flows are using fraction \(n_L/(n_L+n_C)\) of the total link capacity.

We assume that the flow rate of an L4S congestion control is RTT-independent, which is one of the TCP Prague requirements, and has been shown to be feasible in recent work. Then coupled markings purely divide up capacity, and are independent of the queuing latency in the L4S queue. That is, whatever depth queue the L4S AQM is configured for, the coupled marking will not alter capacity shares.

If the native L4S AQM marks packets based on a virtual queue, it will only slightly inflate the marking level so that the L4S queue will be shorter and the L4S flows will fit into \((1-\varepsilon)n_L/(n_L+n_C)\) of the capacity, which is slightly less than with marking based on a real queue. All the flows (Classic and L4S) will then adjust their congestion windows upwards slightly (by about \(\varepsilon\)) to take up the tiny fraction of unused capacity  \(\varepsilon n_L/(n_L+n_C)\), and all the marking and dropping probability will consequently reduce slightly. This slight adjustment should be the only change to capacity shares.

In summary:
\begin{itemize}
	\item Marking based on a virtual queue would be an extremely simple addition to the native L4S AQM;
	\item There should be no need to alter any other part of the DualQ Coupled AQM.
\end{itemize}

However, the simple form of virtual queue measures the the queue in bytes, not in time, which is addressed below.

\subsection{Virtual Sojourn Time}\label{l4saqmtr_vq-sojourn}

As was explained in \S\,\ref{l4saqmtr_current}, L4S marking should be based on a queue measured in units of time. This would be necessary in many cases, even if there were no Classic traffic varying the L4S drain rate\footnote{E.g. in a data centre with only a native L4S traffic in a native L4S AQM without the DualQ or coupling}, because:
\begin{itemize}
	\item The AQM could be part of a larger scheduling hierarchy so traffic in higher priority queues would vary the capacity left for a lower priority queue, rapidly varying the drain rate that the whole DualQ AQM experiences.
	\item Access links often change their rate when modems retrain in response to interference. 
	\item The capacity of radio links varies rapidly over time.
\end{itemize}

\begin{figure}[h]
	\centering
	\includegraphics[width=\columnwidth]{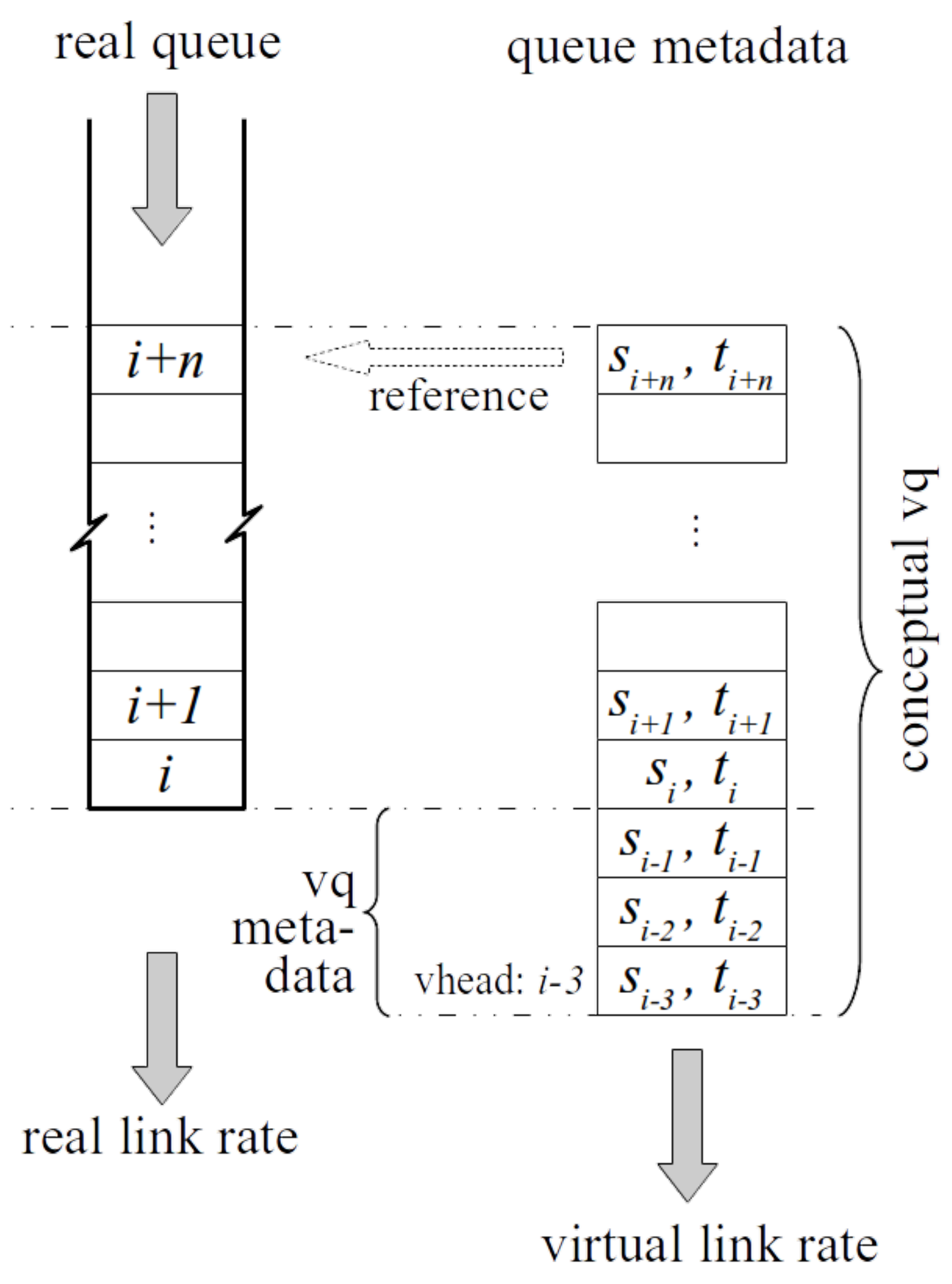}
	\caption{Virtual Sojourn Time}\label{fig:vq-sojourn}
\end{figure}

\autoref{fig:vq-sojourn} shows how an implementation of queue measurement based on sojourn time could be modified to measure the virtual sojourn time through the virtual queue. This would offer an alternative to the variable rate virtual queue patented in~\cite{Eardley14:vary_vq_patent}. The real queue is shown on the left, and the metadata about the packets in the queue is shown on the right, including:
\begin{itemize}
	\item the sizes of each packet (\(s_i\) for the \(i\)th packet);
	\item the timestamps (\(t_i\)) added when each packet was enqueued.
\end{itemize}

\begin{figure*}
\small
\begin{verbatim}
 1. #define LGE 6   /* LGE = lg(1/epsilon) */
 2. #define T 1     /* threshold [ms] */
 3. 
 4. dequeue() {
 5.     now = get_realtime()
 6.     vtsojourn = now - vhead.t           /* Virtual sojourn time */
 7.
 8.     /* Simple illustrative AQM */
 9.     if (vtsojourn > T) {                /* if sojourn exceeds T */
10.         mark head.ecn                   /* mark head real pkt */
11.     }
12. 
13.     s = head.s                          /* size of real head packet */
14.     /* Dequeue real packet (not shown for brevity) */
15.
16.     /* Dequeue virtual packet */
17.     vs = s + s>>LGE                     /* virtual packet size to dequeue */ 
18.     while (vs > vhead.s) {
19.         vs -= vhead.s
20.         /* deallocate metadata of virtual head packet */
21.         vhead++
22.     }
23.     /* reduce size of remaining virtual head packet by remainder */
24.     vhead.s -= vs
25. }
\end{verbatim}
\caption{Virtual Sojourn Time Example Pseudocode}\label{fig:vq-sojourn-code}
\end{figure*}

Everything above the horizontal dashed line drawn across the bottom of the real queue would already be necessary to measure sojourn time and to manage packets in a real queue. Everything below that line implements the virtual queue. Conceptually the virtual queue consists of all the metadata---the real and virtual metadata), as indicated by the brace on the right.

The VQ is essentially created by deferring deletion of some of the metadata about packets that have recently been forwarded, as illustrated by the three example items of metadata shown below the head of the real queue. The variables within the metadata keep the values they had when they referred to real packets, except the size of the head virtual packet (\(s_{i-3}\) in the illustration) is not necessarily as large as the original real packet was.

The only other variable used to maintain the virtual queue is the index of the head packet, \texttt{vhead}, which is shown set to \((i-3)\) in the illustration.

An example implementation of virtual sojourn time is presented in \autoref{fig:vq-sojourn-code}, and a walk-through of the code follows. A very simple time-threshold based AQM is included for illustration, but any other AQM that uses virtual sojourn time could replace it.

Enqueue code is not shown because it is no different to the enqueue code for a real queue that supports sojourn time. It essentially enqueues the packet and adds a metadata structure about the packet at the tail of the queue, which holds the tail packet size, \(s_{i+n}\), and the time, \(t_{i+n}\), when it arrived.

The dequeue code starts by defining macros in lines 1--2 then the dequeue function consists of the following 4 stages:
\begin{description}
	\item[5--6:] Measuring the virtual sojourn time
	\item[8--1:] Illustrative AQM
	\item[13--14:] Dequeuing the real head packet (not shown for brevity
	\item[16--25:] Dequeuing the virtual head packet
\end{description}
The virtual packet is deliberately dequeued second, so that it does not hold up forwarding of the real packet.

Note that code to continue to dequeue the virtual queue as long as the real queue is empty is not shown for brevity.

The following notes walk-through the salient lines:
\begin{description}
	\item[1:] Rather than define \(\varepsilon\) directly, \texttt{LGE} is defined as the base 2 log of the reciprocal of \(\varepsilon\), which is used in a rightwards bit-shift (line 17) that is equivalent to dividing by  \(\varepsilon\).
	\item[2:] Defines the time threshold for the illustrative AQM.
	\item[6:] The timestamp at the head of the virtual queue is subtracted from the current system time to measure the virtual sojourn time.
	\item[9--10:] The real packet is marked if the resulting sojourn time exceeds the marking threshold.
	\item[17:] The size of the virtual packet to dequeue (\texttt{vs}) is calculated from the size of the real head packet, \texttt{s}, inflated by \((1+\varepsilon)\), using a bit-shift for efficiency, as already described.
	\item[18--19:] Because real and virtual packets may all have different sizes, dequeuing \texttt{vs} bytes from the head of the virtual queue might dequeue a number of smaller virtual packets. So this while loop uses \texttt{vs} to hold the remaining virtual bytes to be dequeued. It continues removing virtual head packets and decrementing \texttt{vs} by the bytes removed until \texttt{vs} is less than the size of the next virtual head packet.
	\item[21:] Each time the while loop removes a virtual packet's metadata, it increments the pointer to the head of the virtual queue.
	\item[24:] Finally, it subtracts the remaining value of \texttt{vs} from the new virtual head packet's size. Of course, if \texttt{vs} was intially smaller than the size of the head virtual packet, execution will not enter the while loop at all, and it will pass straight to this line.
	\item[25:] Optionally, the routine accumulates the total count of virtual bytes dequeued in \texttt{vcount\_deq}.
\end{description}

In summary, it can be seen that virtual sojourn time adds very little complexity, because it re-purposes the structures already used for sojourn time in real queues. The while loop to remove multiple small virtual packets seems to be of concern, because it will take an indeterminate time to execute. However, it need not delay forwarding of the real packet, because it is merely tidying up and preparing for the next packet.

If necessary, the execution time of line 17 could be shifted to enqueue by inflating the size of each virtual packet when it is enqueued. However, this is not recommended, because it would require an extra variable in each packet's metadata to store the inflated size, rather than re-using the existing size metadata as shown.

\subsection{Expected Virtual Service Time}\label{l4saqmtr_vq-sojourn-scaled}

\cite{Briscoe17b:sigqdyn_TR} explains that sojourn time fails to take account of evolution of the queue since the head packet arrived at the tail. The same technical report also proposes that a solution would be to scale the sojourn time by the ratio of the backlog at dequeue and at enqueue, that is \texttt{backlog\_deq / backlog\_enq}. For a virtual queue, the ratio between the virtual backlogs would have to be used instead. 

\begin{figure}[h]
	\centering
	\includegraphics[width=\columnwidth]{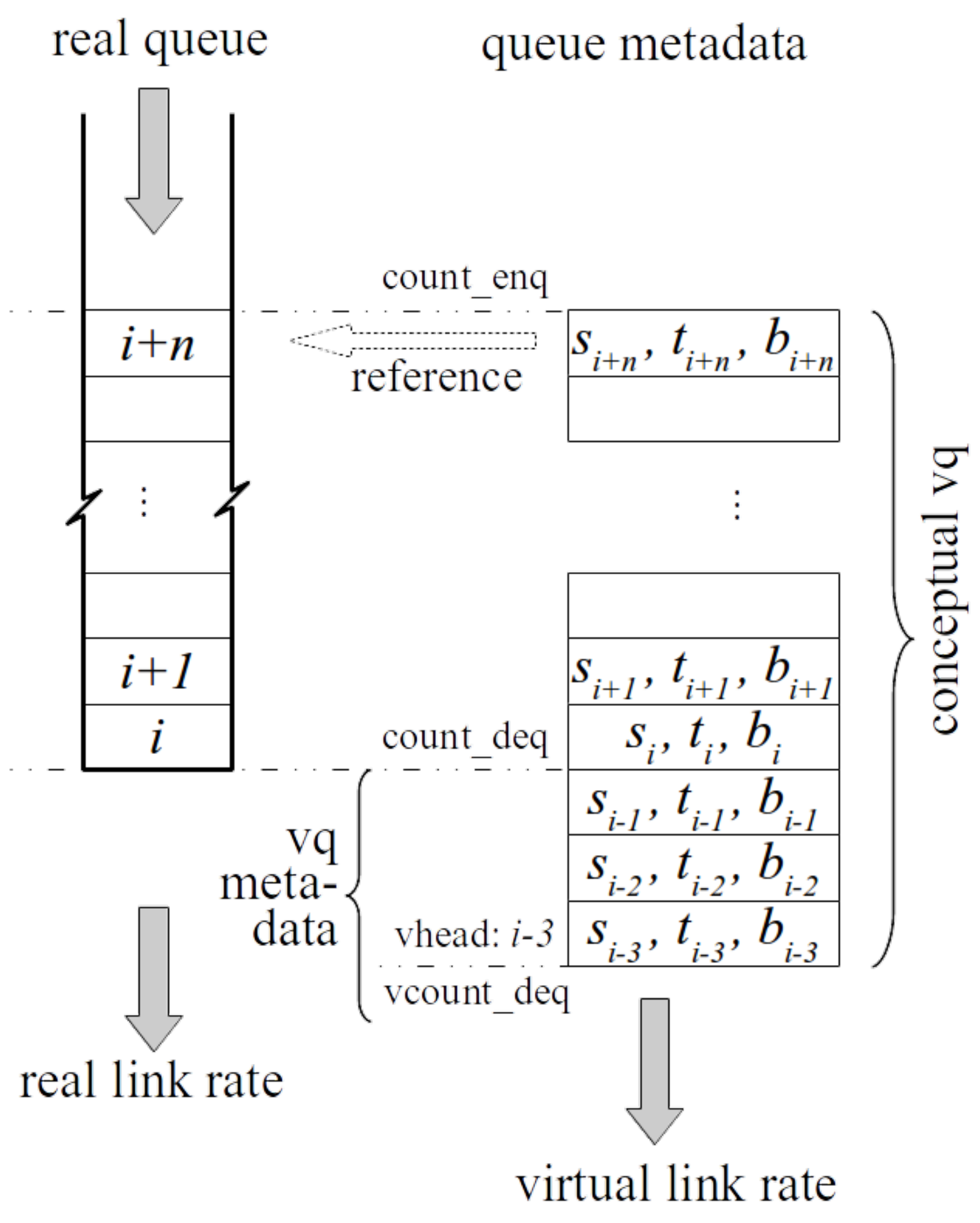}
	\caption{Virtual Sojourn Time}\label{fig:vq-scaled-sojourn}
\end{figure}

\autoref{fig:vq-scaled-sojourn} illustrates the extra variables necessary to implement scaled sojourn time. Accumulated counts of bytes enqueued and dequeued, \texttt{count\_enq} and \texttt{count\_deq}, are maintained by any efficient implementation of a queue. In addition, a count of virtual bytes dequeued, \texttt{vcount\_deq}, would be needed. Also, as noted in \cite{Briscoe17b:sigqdyn_TR}, the virtual backlog at enqueue,  \texttt{vbacklog\_enq}, would need to be calculated  (=\texttt{count\_enq - vcount\_deq}) as each packet was enqueued then stored with the packet's metadata, shown as \(b_i\). 

\begin{figure*}
\small
\begin{verbatim}
 1. #define LGE 6   /* LGE = lg(1/epsilon) */
 2. #define T 1     /* threshold [ms] */
 3. 
 4. dequeue() {
4b.     vbacklog_deq = count_enq - vcount_deq
 5.     now = get_realtime()
 6.     vtsojourn = now - vhead.t           /* Virtual sojourn time */
6b.     /* Scale vtsojourn by ratio of virtual backlogs */
6c.     vtsojourn <<= (clz(vhead.vbacklog_enq) - clz(vbacklog_deq))
 7.
 8.     /* Simple illustrative AQM */
 9.     if (vtsojourn > T) {                /* if sojourn exceeds T */
10.         mark head.ecn                   /* mark head real pkt */
11.     }
12. 
13.     s = head.s                          /* size of real head packet */
14.     /* Dequeue real packet (not shown for brevity) */
15.
16.     /* Dequeue virtual packet */
17.     vs = s + s>>LGE                     /* virtual packet size to dequeue */ 
17b.    vcount_deq += vs
18.     while (vs > vhead.s) {
19.         vs -= vhead.s
20.         /* deallocate metadata of virtual head packet */
21.         vhead++
22.     }
23.     /* reduce size of remaining virtual head packet by remainder */
24.     vhead.s -= vs
25. }
\end{verbatim}
\caption{Virtual Sojourn Time Example Pseudocode}\label{fig:vq-scaled-sojourn-code}
\end{figure*}

\cite{Briscoe17b:sigqdyn_TR} mentions (but does not fully recommend) an implementation trick when scaling sojourn time. it applies the \texttt{clz()} (count leading zeros) function to \texttt{count\_enq}. If this trick were used, it could be applied before storing the value in the metadata, which would reduce the storage requirement to under an 8-bit integer per packet. Alternatively, if the sojourn time were stored as a floating point number, the result could be subtracted from the exponent without any additional storage.

\autoref{fig:vq-scaled-sojourn-code} shows how the implementation of queue measurement based on sojourn time in \autoref{fig:vq-sojourn-code} could be modified to scale the virtual sojourn time. The only modifications are where additional lines 4b, 6b, 6c, and 17b have been inserted. As before, the \texttt{enqueue()} function is not shown.

\section{Gradient Marking}\label{l4saqmtr_gradient}

In DCTCP~\cite{Alizadeh10:DCTCP} and derivate work using a virtual queue, such as HULL~\cite{Alizadeh12:HULL}, the RED AQM that is available in some switch hardware is configured with degenerate values to give a step function. Specifically, RED provides a ramp, but DCTCP sets the min and max threshold of the ramp to the same value, \(K\). RED also uses a smoothed average of the queue, but DCTCP sets the smoothing parameter to zero, in order to use the instantaneous queue.

Using the instantaneous queue seem to be the correct approach, because it removes any smoothing delay before transmitting congestion notification signals. This minimizes delay before the sender receives the signal, then it can either react immediately, or add smoothing itself, after receiving the feedback. Also it can smooth the flows own RTT, whereas the network does not know the RTTs of each flow, so it has to smooth assuming the worst-case RTT (the speed of light in glass round the circumference of the earth, or about 250\,ms).

In \S\,2.2 of \cite{Briscoe17c:l4saqm_TR}, it is argued that randomness should also be shifted from the network to the sender, to place delays due to uncertainty after thye signal has arrived at its destination. A step function satisfies this criterion, but so can other functions.

Using the step function itself is more controversial. The innovation of DCTCP relative to classic TCP was to be able to measure the extent of explicit congestion signalling, not just its existence, within an RTT. However, flows in slow-start or longer-riunning flows in congestion avoidance tend to induce either one whole RTT without marks or one whole RTT with marks. Therefore, in DCTCP's congestion avoidance phase, DCTCP configures the parameter \(g\) to smooth signals over about 16 RTTs. This is because each RTT of signals is effectively either on or off the whole time, with only a little variation due to small flows arriving and departing.

With a gradient marking function, instead of a step, it should be possible for the queue's operating point to sit within the gradient so that a more stable proportion of the packets are marked at a given load, and DCTCP can smooth the signal over one RTT. This would remove a large part of the 16 RTTs of smoothing delay. 

Research is being conducted separately to alter the sending behaviour at the start of a flow to probe for the available capacity much more quickly but with much less harm to other flows and the flow itself than TCP slow-start. Techniques being tried include dithering the intervals between packets, e.g.\ sending different length microbursts at the local line rate, or chirping~\cite{Kuehlewind10b:chirp_impl}.

These techniques cannot rely on the presence of a particular ECN marking regime at the bottleneck (which might be a legacy queue). However, it would be useful to ensure that any ECN marking scheme there is gives such techniques rapid clean information. A step function achieves this at the expense of the on-off behaviour already described. The challenge is to find an alternative that is deterministic, but not on-off, and within the constraints enumerated in the introduction to \cite{Briscoe17c:l4saqm_TR}. 

It would be possible to use a PI controller; for the L4S AQM as well as the base AQM. it could be configured with a very low target for queueing delay (or virtual queuing delay). And randomness could be removed by taking the reciprocal of the output variable p as the interval between marks (or drops). 

Ae PI controller would probably be the best option where a dualQ Coupled AQM was being implemented that already used a PI controller for the base AQM. The PI controller could use time to measure the queue, either using sojourn time, virtual sojourn time or scaled virtual sojourn time.

\subsection{Backwards Compatibility}\label{l4saqmtr_gradient-compat}

For deployment scenarios where only a native L4S AQM was needed (e.g. homogeneous data centres), and therefore implementation changes would not be necessary to deploy the dualQ, it could make sense to use the RED algorithm available in existing hardware. This would only work well for fixed drain-rate single queues, because RED uses queue length, not time as its metric. 

In such deployment scenarios, it is proposed that RED is configured with a steep ramp with \(max_p=1\), with at least half a dozen packets difference in the queue between the minimum and maximum of the ramp, and with no smoothing delay.

Ideally randomness would be removed, as described for PI above. But that would be unlikely to be possible with existing hardware.


\addcontentsline{toc}{section}{References}

{\footnotesize%
\bibliography{aqm-details}}


\onecolumn%
\addcontentsline{toc}{part}{Document history}
\section*{Document history}

\begin{tabular}{|c|c|c|p{3.5in}|}
 \hline
Version &Date &Author &Details of change \\
\hline\hline
01                      &07 Sep 2017   &Bob Briscoe &First complete version.\\\hline%
\metaversion &\metadate     &Bob Briscoe &Added abstract and keywords.\\\hline%
\hline%
\end{tabular}

\end{document}